\documentclass[prd,floatfix,showpacs,eqsecnum,superscriptaddress,nofootinbib]
{revtex4}
\usepackage{epsfig}
\usepackage{amsmath,amssymb}
\usepackage[squaren]{SIunits}

\newcommand{\eq}[1]{Eq.~(\ref{#1})}
\newcommand{\fig}[1]{Fig.~{\ref{#1}}}
\newcommand{\be}{\begin{equation}}
\newcommand{\ee}{\end{equation}}
\newcommand{\bea}{\begin{eqnarray}}
\newcommand{\eea}{\end{eqnarray}}
\newcommand{\DS}{Dyson--Schwinger }
\newcommand{\w}{\omega}

\newcommand{\al}{\alpha}

\newcommand{\ga}{\gamma}

\newcommand{\pd}{\partial}

\newcommand{\cs}{{\cal S}}

\newcommand{\im}{\mathrm{i}}

\begin{document}


\newcommand{\JLU}{Institut f\"ur Theoretische Physik,
  Justus-Liebig-Universit\"at Giessen, Heinrich-Buff-Ring 16, 35392
  Giessen, Germany}   
\newcommand{\TUD}{Theoriezentrum, Institut f\"ur Kernphysik, 
  Technische Universit\"at Darmstadt, 
64289 Darmstadt, Germany}

\title{Fermi velocity renormalization and dynamical gap generation in graphene}
\author{C.~Popovici}
\affiliation{\JLU}

 \author{C.~S.~Fischer}
\affiliation{\JLU}

 \author{L.~von~Smekal}
\affiliation{\JLU}
\affiliation{\TUD}

\begin{abstract}
We study the renormalization of the Fermi velocity by the long-range Coulomb
interactions between the charge carriers in the Dirac-cone approximation for
the effective low-energy description of the electronic excitations in graphene
at half filling. Solving the coupled system of Dyson-Schwinger equations for
the dressing functions in the corresponding fermion propagator with various
approximations for the particle-hole polarization we observe that Fermi
velocity renormalization effects generally lead to a considerable increase of
the critical coupling for dynamical gap generation and charge-density wave
formation at the semimetal-insulator transition.
\end{abstract}

\date\today

\maketitle

\section{Introduction}

Since its synthesis in 2004 \cite{Novoselov2004, Novoselov2005}, graphene has
revealed fascinating electronic properties, such as the anomalous quantum Hall
effect \cite{Gusynin2005,Zhang2005}, Klein tunneling \cite{Katsnelson2006,
  Cheianov2006} or charge confinement \cite{Rozhkov2011} (comprehensive
overviews can be found in
Refs.~\cite{kouc2008,CastroNeto:2009zz,Peres2010,Kotov:2010yh}).  It has been
known for a long time from a simple noninteracting tight-binding model
\cite{Wallace1947,Slon1958} that graphene's twodimensional honeycomb lattice
is special in that its quasi-particle dispersion relation for low energy
excitations around charge neutrality at two independent Dirac points in the
first Brillouin zone is linear, $E_k =\pm v_F |k| $, where $v_F\sim 10^6\,
\mathrm{m/s}\approx c/300 $ is the Fermi velocity and $k = (k_x,k_y)$ is the
momentum of the fermionic quasi-particle in two dimensions. The tight-binding
Hamiltonian effectively reduces to a free Dirac Hamiltonian. This
quasi-relativistic regime for low-energy excitations is separated from the
non-relativistic excitation spectrum at higher energies by a van Hove
singularity in the density of states. Thus already the tight-binding model
also serves as a simple example of an excited state phase transition
\cite{Dietz:2013sga} which in this case reflects the presence of a topological
electronic ground-state transition when the chemical potential is tuned away
from half filling and beyond the van Hove singularity, where the static
susceptibility diverges logarithmically indicative of a neck-disrupting
Lifshitz transition in two dimensions.

Within field theory descriptions of the electronic excitations in graphene,
charge fractionalization and vortex formation have been described through
chiral gauge models \cite{Hou:2006qc,Jackiw2007,Oliveira:2010hq} which include
the dynamics of the carbon background and doping effects. Charge confinement
and Klein tunneling were investigated within such a model in
\cite{Popovici:2012xs}. Cosmological models were used to describe the
electronic properties of deformed graphene sheets \cite{Cortijo:2006xs},
topological defects and curvature as described by geometric gauge fields also
lead to an index theorem for graphene \cite{Pachos:2006ah}. The various uses
of gauge fields to model topological defects as well as stress and strain on
graphene sheets are reviewed in \cite{Vozmediano:2010zz}.  The effects of
additional short-range four-fermion couplings to model phonon interactions
have been investigated at mean-field level in \cite{Gamayun:2009em}, and more
recently within the Functional Renormalization Group approach in
\cite{Mesterhazy:2012ei,Janssen:2012pq}.

The importance of many-body interactions in graphene has been established both
theoretically (see Ref.~\cite{CastroNeto:2009zz} and references therein, for
example) and experimentally (in a recent measurement of the Fermi velocity at
low energies \cite{elgo2012}, but also from the observation of a gap in an
ARPES measurement of epitaxial graphene, upon dosing with small amounts of
atomic hydrogen \cite{boche2009}).  Long-range electron-electron interactions
induce a momentum dependent renormalization of the Fermi velocity. If the
Coulomb interaction can be made sufficiently strong, one furthermore expects
that a mass gap is dynamically generated, analogous to chiral symmetry
breaking in QCD, such that graphene undergoes a phase transition from a
semimetal to an insulator by charge-density-wave formation. This expectation
is motivated by the potentially large value of the effective `fine structure
constant' $\al=e^2/(4\pi\hbar v_F \varepsilon) \sim O(1)$ when the dielectric
constant $\varepsilon$ is of the same order as one would expect for suspended
graphene with $\varepsilon =1 $.

The critical coupling for this semimetal-insulator transition has been
estimated theoretically in a variety of ways generally yielding values of
$\alpha_c \sim 1$ when no additional screening effects on the Coulomb
interactions between the electrons in the graphene $\pi$ bands, {\it e.g.},
from electrons in the $\sigma$ band, are included.  A value of $\alpha_c=0.92$
was obtained in \cite{Gamayun:2009em} from a bifurcation analysis of a
simplified gap equation in which radiative corrections to the Fermi velocity
were neglected. A particular form for a momentum dependent Fermi velocity
renormalization based on a large $N$ expansion was used in \cite{khve2009} to
obtain $\alpha_c=1.13$.  A renormalization group calculation at two-loop order
yielded $\alpha_c=0.833$ \cite{vafek2008}. Lattice studies have reported from
Monte-Carlo simulations on both, standard square \cite{drula2009} and physical
hexagonal \cite{Buividovich:2012nx} lattices $\al_c=1.08\pm 0.05$ and
$\al_c=0.9\pm 0.2$, respectively. All these values are well below the bare
coupling constant of suspended graphene, $\al_0=2.19$ with $\varepsilon=1$,
and thus in apparent contradiction with the experimental observation that
suspended graphene remains in the semimetal phase
\cite{Kotov:2010yh,elgo2012}.

One possible explanation of this discrepancy might be the additional screening
of the Coulomb interaction due to the other electrons, notably those in the
$\sigma$ band. In a constrained random phase approximation these were indeed
found to reduce the on-site repulsion and the first three nearest-neighbor
Coulomb interaction parameters by successively decreasing factors between 1.8
and 1.3 \cite{Wehling:2011}. In a recent Monte-Carlo simulation
\cite{Ulybyshev:2013swa}, these parameters were used together with a
corresponding screening of the long-range Coulomb tails to effectively reduce
the interaction strength by an amount which was found to be sufficient to
place suspended graphene in the semimetal phase. While the additional
screening of the short-range part of the $\pi$-band electrons' Coulomb
interactions should certainly be included in a more quantitative calculation,
here we focus on qualitative effects of their long-range Coulomb tails which
one would in fact not expect to be screened by the other, localized electron
states. A realistic description will probably have to include the right
balance of all effects, the correct amount of on-site repulsion, favoring an
antiferromagnetic Mott-state \cite{Sorella:1992}, the screening of the nearest
neighbor and short range interactions which would otherwise lead to
charge-density-wave formation \cite{herbut2006,Honerkamp:2008}, as well as
unscreened long-range Coulomb tails which might enhance the latter. Even if
the strengths of these competing effects are all too weak for an insulating
ground state in suspended graphene, it will be interesting to find out whether
it is at least close to a possible transition into any one of the insulating
phases (gapped spin-liquid \cite{Meng:2010} and topological insulator phases
\cite{Raghu:2008} have also been discussed).

Whatever the final answer will be, the experimentally observed reshaping of
the Dirac cones in suspended graphene \cite{Kotov:2010yh,elgo2012} does
indicate that the Coulomb interactions induce a charge-carrier density
dependent renormalization of the Fermi velocity. This renormalization peaks at
half-filling where it leads to an increase by about a factor of three in the
Fermi velocity and thus a corresponding decrease in the renormalized effective
coupling. A momentum dependence of the Fermi velocity of a similar kind was in
fact already predicted in an early perturbative RG study \cite{Gonzalez94},
long before the synthesis of graphene. There, it was even concluded that $v_F$
should keep growing logarithmically until retardation effects become important
enough to invalidate the instantaneous Coulomb approximation.

Lattice simulations certainly have the potential to provide reliable results
also for strongly coupled condensed matter systems, if the bare interactions
are chosen correctly to describe the physical system. Even if the microscopic
theory is completely specified as in QCD, however, continuum (at least in the
time discretization), infinite volume and chiral extrapolations are often
difficult and expensive. This is very much so for the simulations of the
electronic properties of graphene also. Moreover, a chemical potential for
charge-carrier densities away from half filling, {\it e.g.}, to study the
effects of interactions on the electronic Lifshitz transition of the free
tight-binding model \cite{Dietz:2013sga}, introduces a fermion sign problem
here as well. Nonperturbative continuum approaches such as the Functional
Renormalization Group or Dyson-Schwinger equations therefore provide valuable
alternatives. Especially when the necessary truncations are tested against
lattice simulations in finite systems, without massless fermions and sign
problem, the corresponding limits as well as finite density effects are
relatively straightforwardly addressed with these functional continuum
methods. In this paper we employ the \DS approach, which has been successfully
applied to both QCD and three-dimensional QED (see, for example,
Refs.~\cite{Alkofer:2000wg,Maris:2003vk,Fischer:2006ub,Maris:1995ns,
fial2004,Bonnet:2011hh,Bonnet:2012az,Popovici:2013fya}
and references therein). In particular, we study the running of the Fermi
velocity from Coulomb interactions in the Dirac-cone approximation, and its
influence on the critical coupling for the semimetal-insulator transition by
charge-density-wave formation in the effective low-energy model. To this end,
we extend the study of Ref.~\cite{Gamayun:2009em} by first calculating the
running Fermi velocity in various approximations for the particle-hole
polarization. The results are qualitatively in line with the observed
reshaping of the Dirac cones. In order to then determine the corresponding
values for the critical coupling we numerically solve the coupled system of
\DS equations for the fermion propagator with and without running Fermi
velocity for comparison. We generally observe that the resulting critical
coupling substantially increases when the strong renormalization effects in
the Fermi velocity are included, thus favoring the persistence of suspended
graphene in the semimetal phase. This general trend has also been observed in
our preliminary analysis \cite{Popovici:2013fwp} in which we employed a GW
approximation, in addition, in order to compute the Fermi velocity
renormalization analytically. 

The paper is organized as follows: In the next section we briefly review the
continuum model of graphene which is a variant of QED$_3$ with instantaneous
bare Coulomb interactions. In Section \ref{sec3} we discuss the
Dyson-Schwinger equation (DSE) for the fermion propagator in this model and
present our solutions for the different particle-hole polarizations. Thereby,
we first describe the results from a bifurcation analysis to find the critical
point in Subsec.~\ref{sec3A}.  Our iterative numerical solutions of the full
system of coupled integral equations are then presented in
Subsec.~\ref{sec3B}. Thereby we verify the bifurcation analysis and obtain
complete results for the fermion mass and Fermi velocity renormalization
functions in both phases.  We compare our results for $\al_c$ to the
literature on the DSE approach and assess the validity of the various
approximation schemes.  Moreover, we discuss the behavior of the
pseudocritical coupling with explicit symmetry breaking by a staggered
chemical potential, and the dependence of the critical coupling in the chiral
limit on the number of fermion flavors providing evidence of Miranski scaling.
Our summary and outlook are provided in Section \ref{sec5}.


\section{Details of the model}
\setcounter{equation}{0}

In this paper we study a low-energy continuum model for the long-range Coulomb
interactions between the charge carriers in monolayer graphene, as previously
considered in
Refs.~\cite{gonzalez1999,khve2001,gogu2001,gogu2002,Gamayun:2009em}.  In this
model, the quasi-particle excitations at energies well below the van Hove
singularity, around the two Dirac points within the first Brillouin zone of
the honeycomb lattice, are described by massless Dirac fermions in two spatial
dimensions. The honeycomb lattice is built from two independent triangular
sublattices, corresponding to a triangular Bravais lattice with a
two-component basis. Consequently, four-component Dirac spinors are introduced
for the quasi-particle excitations on both sublattices $A$ and $B$, each with
momenta close to either of the two inequivalent Dirac points $K$ (plus sign)
and $K^\prime$ (minus sign), $\psi^{T} =\left(\psi^B_{+s},\, \psi^A_ {+s},\,
\psi^A_{-s} ,\, \psi^B_{-s} \right)$. The true spin of the electrons formally
appears as an additional flavor quantum number $s=1,...,N_f$ with $N_f=2$ for
monolayer graphene. The graphene model is then specified by the action (in
natural units with $\hbar=c=1$)
\be
\cs=\int dt\, d^2r\,\, \bar\psi(t,\vec r)
\left[\im\ga^0\pd_t-\im v_F\ga^i\pd_i\right]\psi(t,\vec r)
+\cs_{int} \, , \label{eq:origaction}
\ee
where $v_F \approx 1/300 $ is the Fermi velocity (we will return to its
definition in the next section), $\psi$ and $\bar\psi=\psi^\dagger\ga^0$
denote the pseudo-spin $1/2$ fermion field and its Dirac adjoint. The spatial
index is $i= 1,2$, and the three 4-dimensional $\gamma$-matrices reducibly
represent the Clifford algebra $\{\ga^{\mu},\ga^{\nu}\}=2g^{\mu\nu}$ in 2+1
dimensions. In absence of magnetic fields and with $v_F\ll 1$, the
electromagnetic interactions reduce to the purely electric coupling with the
zero-component of the photon field, initially in 3+1 dimensions (with $\vec
x=(\vec r, z)$) and Feynman gauge,
\be
\cs_{int} =\int dt\, d^3x\left[-\rho(t,\vec x)A^0(t,\vec x)
  +\frac{1}{2}\big(\vec\nabla A^0(t,\vec x)\big)^2
  -\frac{1}{2}\big(\partial_0 A^0(t,\vec x)\big)^2  \right]\, , 
\mathrm{~~~with~~}
\rho(t,\vec x)=e \bar\psi(t,\vec r) \ga^0 \psi(t,\vec r) \delta(z)\, .
\label{eq:intac}
\ee
Magnetic interactions with the spatial vector components of the photon field
could be introduced via Peierls substitution but are $O(v_F)$ suppressed and
hence neglected. Consequently, the bare photon propagator in the $z=0$ plane
is simply given by
\be
D_\mathrm{tl}^{00}(t,\vec r,z=0) = \int \frac{d\omega}{2\pi} \frac{d^2k}{(2\pi)^2}
\frac{dk_z}{2\pi} \,  \frac{-\im}{\omega^2-\vec k^2 -k_z^2 +
  \im\epsilon} \; e^{-\im\omega t}  e^{\im \vec k\vec r} \, .
\ee
Integration over the perpendicular $k_z$-momentum modes of the Coulomb photon
in the instantaneous approximation \cite{Gonzalez94} then yields the
dimensionally reduced, so-called ``brane action'' for the quasi-particles
\cite{gogu2001} with Coulomb interactions as usually employed in condensed
matter systems,
\be
\cs=\int dt\, d^2r\,\, \bar\psi(t,\vec r)
\left[\im\ga^0\pd_t-\im v_F\ga^i\pd_i\right]\psi(t,\vec r)
-\frac{e^2}{8\pi\varepsilon}\int dt \, d^2r \, d^2r^{\prime}\, 
\bar\psi(t,\vec r)\ga^0\psi(t,\vec r) \frac{1}{|\vec r-\vec
  r^{\,\prime}|}
\bar\psi(t,\vec r^{\,\prime})\ga^0\psi(t,\vec r^{\,\prime})\, .
\label{eq:braneaction}
\ee
Here we have also introduced a dielectric constant $\varepsilon = (1+\kappa)/2
$ to model the screening of the Coulomb interactions on top of a substrate,
typically with $\kappa \approx 4 $ for SiO$_2$ or $\kappa\approx 10 $ for SiC,
instead of $\varepsilon=1$ for suspended graphene (neglecting the additional
short-range screening from the localized electron states in graphene here,
which would require a momentum dependent $\varepsilon(\vec k)$
\cite{Wehling:2011}).  As discussed below, the bare Coulomb propagator is
further modified due to the interactions by the particle-hole polarizability
$\Pi(\omega,\vec k)$, or Lindhard function of the electrons in the
$\pi$-bands. 

The Feynman rules for the model are specified by the tree-level electron
propagator and the fermion-photon vertex.  These can be read off from the
action as specified by Eqs.~(\ref{eq:origaction}) and (\ref{eq:intac}) to have
the usual form with dimensionless charge despite the reduced dimensionality of
the brane action, 
\bea
&&S_\mathrm{tl}^{-1}(p_0,\vec p)=-\im(\ga^0p_0-v_F\ga^ip_i)\, ,\\
&&\Gamma_\mathrm{tl}^{0}(k,p)=-\im e\ga^0\, .
\eea
In addition to the tree-level quantities, we will also require the general
decomposition for the fermion propagator $S(p)$.  Because of the anisotropy
introduced by the Fermi velocity, we need to treat temporal and spatial
components separately,
\be
S(p)=\frac{\im ( Z(p)\,\ga^0p_0-v_F A(p)\,\ga^ip_i+\Delta(p))}
{Z^{2}(p)\,p_0^2-v_F^2A ^2(p)\,\vec p^{\,2}-\Delta^2(p)+\im\epsilon}\, ,
\ee
where $Z(p)$ is the wavefunction renormalization, $A(p)$ is the Fermi velocity
dressing function, and $\Delta(p)$ is the quasi-particle gap or mass
function. These quantities can be obtained by solving numerically the \DS
equation for the fermion propagator.  The determination of the critical point,
along with the corrections introduced by the Fermi velocity renormalization,
will be discussed in the next section.

\section{The \DS equations}\label{sec3}
\setcounter{equation}{0}

Starting from the action in \eq{eq:braneaction}, the fermion \DS (gap)
equation follows with standard techniques,
\be
S^{-1}(p_0,\vec p)=S_\mathrm{tl}^{-1}(p_0,\vec p)+\im e \int
\frac{d^3k}{(2\pi)^3} \; \ga^0 S(k_0,\vec k)\Gamma^0(k,p)
D_C(p_0-k_0,\vec p-\vec k)\,,   \label{fDSE}
\ee
where $\Gamma^0(k,p)$ is the fully dressed fermion-photon vertex depending on
the incoming and outgoing fermion momenta.  $D_C(q_0,\vec q)$ is the
dimensionally reduced, instantaneous Coulomb propagator with
frequency-dependent Lindhard screening via the inclusion of the collective
particle-hole polarizability.  In the random phase approximation (RPA), with
the one-loop expression for the polarization function $\Pi(q_0,\vec q)$, it is
given by \cite{gonzalez1999}
\be\label{one-loop}
D_C(q_0,\vec q)=\frac{\im}{2 |\vec q|+\Pi(q_0,\vec q)}\, 
, \;\; \mathrm{~~with~~}\quad \Pi(q_0,\vec
q)=\frac{e^2 N_f}{8\varepsilon}\frac{\vec q^{\,2}}{\sqrt{v_F^2\vec
    q^2-q_0^2}}\, . 
\ee
Note, however, that by definition this one-loop expression for $\Pi(q_0,\vec
q)$ does not include radiative corrections to the Fermi velocity due to the
renormalization function $A(p)$. As we will demonstrate explicitly below,
strong Fermi velocity renormalization effects indeed tend to suppress the
Lindhard screening, whose effect on the bare Coulomb interaction is hence
overestimated by the one-loop polarizability $\Pi(q_0,\vec q)$ in
\eq{one-loop}. In the static limit, $q_0 \to 0$, the fully instantaneous RPA
Coulomb propagator simply reduces to the bare one with constant Lindhard
screening,  
\be\label{instphoton}
\Pi(0,\vec q)=\frac{e^2 N_f}{8\varepsilon}\frac{|\vec{q}|}{v_F} =
2 g |\vec{q}| \, , \;\; \mbox{and} \quad  D_C(0,\vec q) = \frac{1}{1+g}\, 
\frac{\im}{2|\vec q|}\, , 
\ee
where $g=\al_\varepsilon N_f \pi/4$ and $\al_\varepsilon =e^2/(4\pi v_F
\varepsilon)$.

We now proceed with Dirac projection and Wick rotation to Euclidian space
($k_0\to \im \w$).  With the fully instantaneous Coulomb interaction, both the
Fermi velocity and mass renormalization functions remain frequency
independent, $A\equiv A(\vec p)$ and $\Delta\equiv \Delta(\vec p)$. One then
furthermore has $Z=1$ and hence, as a consequence of the Ward-Takahashi
identity, also $\Gamma^0 = \Gamma^0_\mathrm{tl}$. As long as the frequency
dependence in the Lindhard screening remains weak, one may therefore assume
that, as good first approximation, wave function renormalization, vertex
corrections and the frequency dependences in $A$ and $\Delta$ can still be
neglected. In this approximation, setting $p_0=0$ in the DSE (\ref{fDSE}), one
obtains the following system of coupled integral equations for the fermion
dressing functions $A(\vec p)$ and $\Delta(\vec p)$ \cite{Gamayun:2009em},
\bea
A(\vec p)
&=&1+\alpha_\varepsilon v_F\int_{-\infty}^{\infty}\frac{d\w}{2\pi}\int\frac{d^2\vec
  k}{(2\pi)^2}
\frac{\vec p\cdot \vec k}{\vec p^{\,2}}\frac{A(\vec k)}{\w^2+v_F^2
  A^2(\vec k)\vec
  k^2+\Delta^2(\vec k)} \big(\!-4\pi \im \, D_C(\im \w, \vec p-\vec
k)\big)\, ,\label{eq:gapA}\\ 
\Delta(\vec p)
&=&\alpha_\varepsilon v_F \int_{-\infty}^{\infty}\frac{d\w}{2\pi}\int\frac{d^2\vec
  k}{(2\pi)^2}
\frac{\Delta(\vec k)}{\w^2+v_F^2 A^2(\vec k)\vec
  k^2+\Delta^2(\vec k)} \big(\!-4\pi \im \, D_C(\im \w, \vec p-\vec
k)\big)\, .\label{eq:gapD}
\eea 
The temporal integration with the RPA Coulomb propagator in \eq{one-loop},
\be
I(\vec p,\vec k)=
\int_{-\infty}^{\infty}\frac{d\w}{2\pi} \; 
\frac{-4\pi \im \,  D_C(\im \w, \vec p-\vec
k)}{\w^2+v_F^2\vec k^2 A(\vec k)^2+\Delta(\vec k)^2} \, ,
\ee 
has already been performed in \cite{Gamayun:2009em}. It is given
by\footnote{Note the explicit inclusion of the bare Fermi velocity $v_f$ here,
  as compared to Ref.~\cite{Gamayun:2009em} in which $v_F = 1$ was used in
  both, the fermion DSE (\ref{fDSE}) and the RPA Coulomb propagator
  (\ref{one-loop}).}
\be
I(\vec p,\vec k)=\frac{J(z,g)}{|\vec p-\vec k| \sqrt{v_F^2\vec
    k^2A^2+\Delta^2}}\, ,
\,\,\,\mathrm{~~with~~} 
z=\frac{\sqrt{\vec k^2 A^2+\Delta^2/v_F^2}}{|\vec p-\vec k|} \, ,
\label{eq:Iz}
\ee
and $J(z,g)$ as a piecewise-defined function,
\bea
J(z,g)=\frac{(z^2-1)[\pi-g c(z)]+zg^2 c(g)}{z^2+g^2-1},
\mathrm{~~with~~}
c(x)=\left\{ 
  \begin{array}{l l  l}
2\,\mathrm{arccosh}(x)/\sqrt{x^2-1} & \quad x>1,\vspace{1mm}\\
2\arccos (x)/\sqrt{1-x^2} & \quad x<1\vspace{1mm},\\
2 & \quad x=1.
  \end{array} \right.
\label{eq:Jc}
\eea
With bare ($\Pi=0$) and fully instantaneous RPA ($\Pi = 2g |\vec q|$, {\it
  c.f.}, \eq{instphoton}) Coulomb interactions, one has 
\be
J(z,g) = \pi\,  \;\; \mbox{and} \quad J(z,g) = \frac{\pi}{1+g}\,, 
\label{eq:Jc2}
\ee
respectively. In both these cases $J$ is independent of the fermion dressing
functions $A$ and $\Delta$. We will also use these two simple special cases in
our analysis of the critical coupling for comparison below.

\subsection{Critical point analysis}
\label{sec3A}

For studying the dynamics at the critical point, an appropriate mathematical
tool is bifurcation theory \cite{atbl1994}. In this framework, the nonlinear
equations simplify such that the critical coupling $\alpha_c$, where the
nontrivial solution of the gap equation bifurcates away from the trivial one,
can be evaluated.  Explicitly, applying bifurcation theory amounts to
expanding Eqs.~(\ref{eq:gapA}) and (\ref{eq:gapD}) to leading order
$\Delta$. This means using $\Delta = 0 $ in (\ref{eq:gapA}), {\it i.e.}, to
solve for $A$ in the symmetric phase, and expanding the right hand side of
Eq.~(\ref{eq:gapD}) to linear order in $\Delta$. In the Dirac cone
approximation we furthermore have rotational invariance, and we can hence
write $A(\vec p)= A(p)$ and $\Delta(\vec p)=\Delta(p)$ from now on with the
notation $p\equiv |\vec p|$ for the magnitude of the spatial quasi-particle
momentum,
\bea
A(p)&=&1+\frac{\alpha_\varepsilon
}{2\pi^2}\frac{1}{p}\int_{0}^{\Lambda} dk\, k  \int_{0}^{\pi} 
d\theta \frac{\cos \theta}{\sqrt{p^2+k^2-2\, p \,k \, \cos\theta}}
J(z_0,g)\, ,\label{eq:gapA2}\\
\Delta(p)&=&\frac{\alpha_\varepsilon}{2\pi^2}\int_{0}^{\Lambda} dk \int_{0}^{\pi}
d\theta \frac{1}{\sqrt{p^2+k^2-2\, p\, k\, \cos\theta}}
\frac{\Delta(k)}{A(k)}J(z_0,g)\, . \label{eq:gapD2}
\eea
Here, $J(z,g)$ is either given by Eq.~(\ref{eq:Jc}) and now $z_0= A(k)\, k/
|\vec p- \vec k|$ for $\Delta = 0$, or simply with either of the two constant
values in (\ref{eq:Jc2}). In the symmetric phase up to and, at a continuous
transition, including the critical point the integral equation
(\ref{eq:gapA2}) for the Fermi velocity renormalization $A$ can be solved
independently. In the broken phase, this solution for $\Delta = 0$ will of
course not be the thermodynamically favored one. For the bifurcation analysis
it can however be inserted into \eq{eq:gapD2}, in order to evaluate the
critical coupling numerically, by a variant of the so-called power method
\cite{Hoffman2001}. 

We emphasize that we solve all angular integrals
numerically thus avoiding further angle approximations as used in many
previous studies. Moreover, it is also worth mentioning that the linearized
equations can be solved much faster than the full nonlinear ones, because the
critical slowing down observed for the latter ones as the critical coupling is
approached does not occur in the linearized problem. In order to
obtain sufficiently precise values for $\alpha_c$ it is crucial to use
an extremely low momentum cutoff in the infrared. 
It has been shown in \cite{Gusynin:2003ww} and \cite{Goecke:2008zh} that
the critical number of fermion flavors in ordinary QED$_3$, for
example, is very sensitive to finite volume effects and thus to the
infrared cutoff of the theory. We observe a similar  
phenomenon in the graphene version of QED$_3$ considered here. In our numerical
integrals we need infrared cutoffs of the order of $p^2 \sim 10^{-10}
\Lambda^2 $ to obtain reliable values of $\alpha_c$, with controllable
systematics. In the $\Pi=0$
case we could furthermore compare our numerical calculation with  an  
analytical result  of 
 $\alpha_c^{\Pi=0} = 8 \pi^2/\Gamma^4(1/4) \approx 0.457$ 
by Gusynin and collaborators \cite{Gusynin:unp,Gamayun:2009a} which 
then yielded very good agreement, with errors below the one percent
level, see below. 

The resulting values of the critical couplings for chiral symmetry breaking
with $N_f=2$ for monolayer graphene in the various approximations are
summarized in Table \ref{tab:alpha}, the error bars reflect our numerical error
estimated from comparison with our results from solving the set of 
non-linear equations detailed below.
\begin{table}[b]
  \centering
  \begin{tabular}{l||c|c|c}
  $\alpha_c$& $\Pi=0$	& $\Pi_{inst.}$	& $\Pi_{one-loop}$	\\\hline
  $A=1$		&  0.457(3)	& 1.63(2)		& 0.81(1)			\\
  $A(p)$	&  1.21(1)	& $\infty$		& 7.65(5)			\\\hline
  \end{tabular}
\hskip .6cm
  \parbox[c]{0.4\linewidth}{\caption{Critical couplings in various
      approximations for the particle-hole polarizability $\Pi$ with and
      without Fermi velocity renormalization $A(p)$.\label{tab:alpha}}}
\end{table}
Compared are the results from the three different approximations for the
particle-hole polarization. For each of them we furthermore compare the
solutions with selfconsistent $A(p)$ and without Fermi velocity
renormalization, {\it i.e.}, $A=1$. The difference between the upper and the
lower row in the table thus serves as a measure of the importance of including
the nontrivial Fermi velocity renormalization function $A(p)$. Clearly, the
effects are huge. Whereas for $\Pi=0$ the critical coupling increases by a
factor of about 2.5, there is no finite solution in the fully instantaneous
approximation at all when $A(p)$ is dynamically included. This can be verified
analytically because the instantaneous and bare Coulomb couplings $\alpha_0$
and $\alpha_{inst.} $ from Eqs.~(\ref{eq:Jc2}) are simply related by
\be
\alpha_{inst.}(\alpha_0)  = \frac{2\alpha_0}{2- \pi \alpha_0} \, . 
\ee
Hence $\alpha_{inst.}(0.457) = 1.62$, but
$\alpha_{inst.} $ diverges at $\alpha_0 = 2/\pi \approx 0.64$ and there is no
positive value for $\alpha_{inst.} $ above that, in particular at $\alpha_0 =
1.21$ with Fermi velocity renormalization. This is in agreement with our
numerical analysis which fails to find a bifurcation point in this case. With
the full RPA Coulomb interaction in (\ref{one-loop}) we obtain $\al_c\approx
0.81$ for $A=1$ as compared to $\al_c\approx 0.92$ obtained in
Ref.~\cite{Gamayun:2009em} from an analogous computation but with an
additional angle approximation.  Again this value dramatically increases to
$\al_c\approx 7.65$ when the selfconsistent Fermi velocity renormalization
$A(p)$ is included.  The general trends can be understood as follows: static
Lindhard screening increases the critical coupling too much because it clearly
overestimates the screening effects by the particle-hole polarizability. These
effects are somewhat reduced by including the frequency dependence in the
Lindhard screening with the full RPA polarization. The strong Fermi velocity
renormalization observed in all cases effectively weakens the Coulomb
interaction and thus increases the critical coupling again. At the same time,
however, one would expect that large deviations of $A(p)$ from $A=1$ would
effectively reduce the particle-hole polarizability again in a fully
selfconsistent solution of the fermion DSE together with that for the
particle-hole polarization function $\Pi$, which would clearly be an important
effect beyond RPA. In such a fully dynamic computation, one might expect two
competing effects, a very large Fermi velocity renormalization would on one
hand effectively switch off the Lindhard screening and hence tend to restore
the bare Coulomb interactions. This, on the other hand, would in turn reduce
the Fermi velocity renormalization to some degree again, not entirely though
because it is still rather strong in our $\Pi=0$ computation. Therefore, one
might expect that the fully dynamical result should be somewhere between
$\al_c\approx 1.21$ for $\Pi=0 $ and $\al_c\approx 7.65$ with the largely
overestimated Lindhard screening in the RPA Coulomb interaction
(\ref{one-loop}), when strong Fermi velocity renormalization effects occur. 

In all approximations we do observe a significant increase of $A(p)$ in the
vicinity of the corresponding critical $\al_c$ such that $A(p)$ is
significantly larger than one in the relevant momentum regime (see below). If
plugged into the particle-hole polarization function such large values of $A$
would drastically suppress the polarization effects in the low momentum
region. As a result, the simple approximation $\Pi=0$ might even be closer to
the correct answer than the one-loop expression in \eq{one-loop}. The final
answer will have to await a fully dynamical and selfconsistent simultaneous
solution to both, the fermion DSE and the particle-hole polarization function
({\it i.e.}, the Coulomb photon DSE).

The various results from Monte-Carlo simulations within lattice gauge theory
mentioned in the introduction are all somewhat below our lower bound.
Especially those by Drut and L\"ahde who obtained $\al_c=1.08\pm 0.05$
\cite{drula2009} should be comparable to our results because they essentially
use a discretized version of the same Dirac-cone approximation as the
effective description for the low-energy electronic excitations of graphene,
however with full QED interactions. Quite obviously, retardation effects might
become important with strong Fermi velocity renormalization
\cite{Gonzalez94}. These are beyond the instantaneous Coulomb interactions
with frequency dependent but non-relativistic screening used here.  For a more
precise comparison between Dyson-Schwinger and lattice results we should first
include the photon polarization dynamically and selfconsistently,
however. Finite volume and finite mass effects could then be analyzed in more
detail also in the DSE approach, in order to be used for infinite volume and
chiral extrapolations.\footnote{For a corresponding comparison in 'ordinary'
 QED$_3$ see e.g. Ref.\cite{Gusynin:2003ww,Goecke:2008zh}.}

\subsection{Nonlinear equations}
\label{sec3B}

\begin{figure}[t]
\vspace{0.5cm}
\centering
\includegraphics[width=0.48\linewidth]{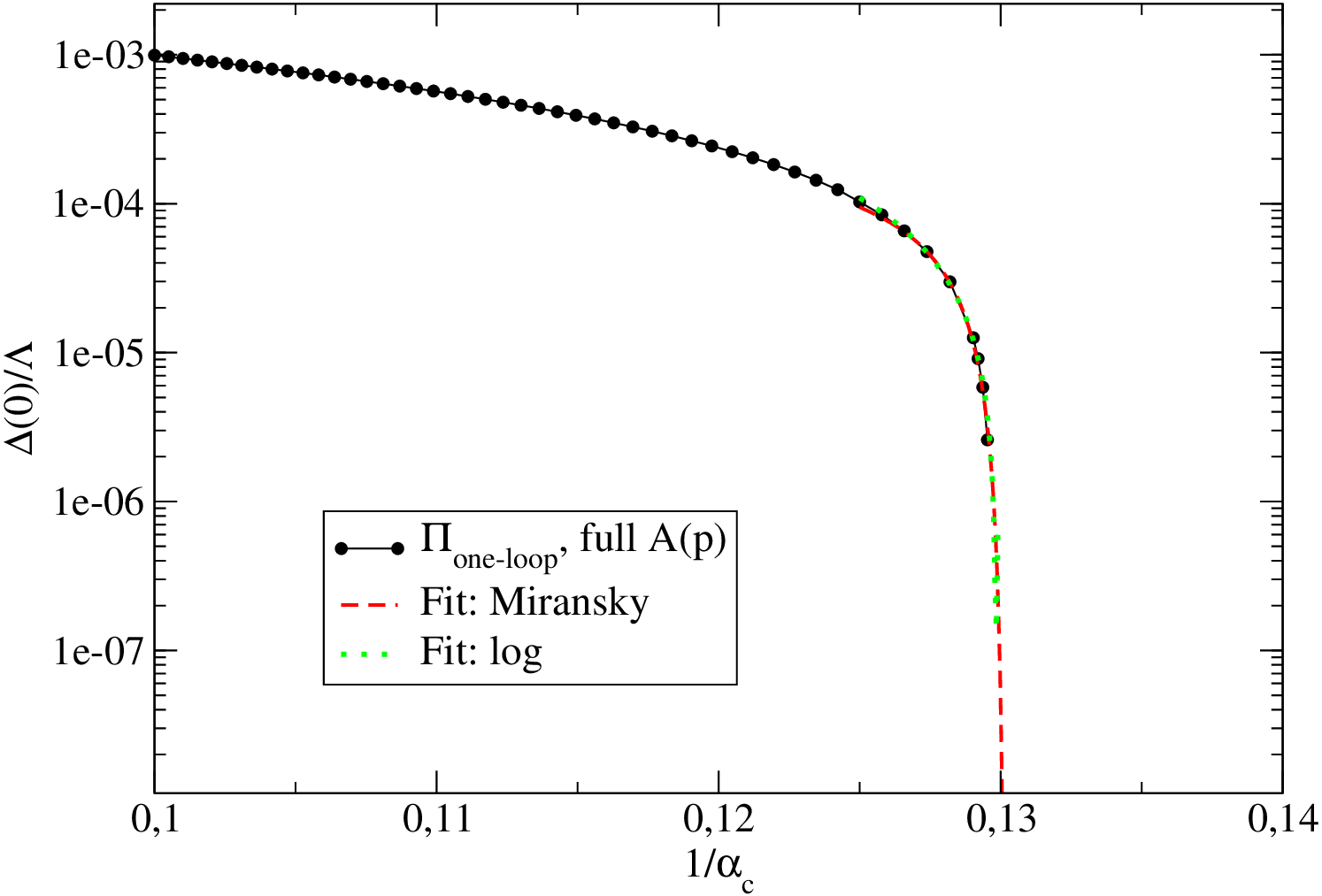}
\hfill
\includegraphics[width=0.44\linewidth]{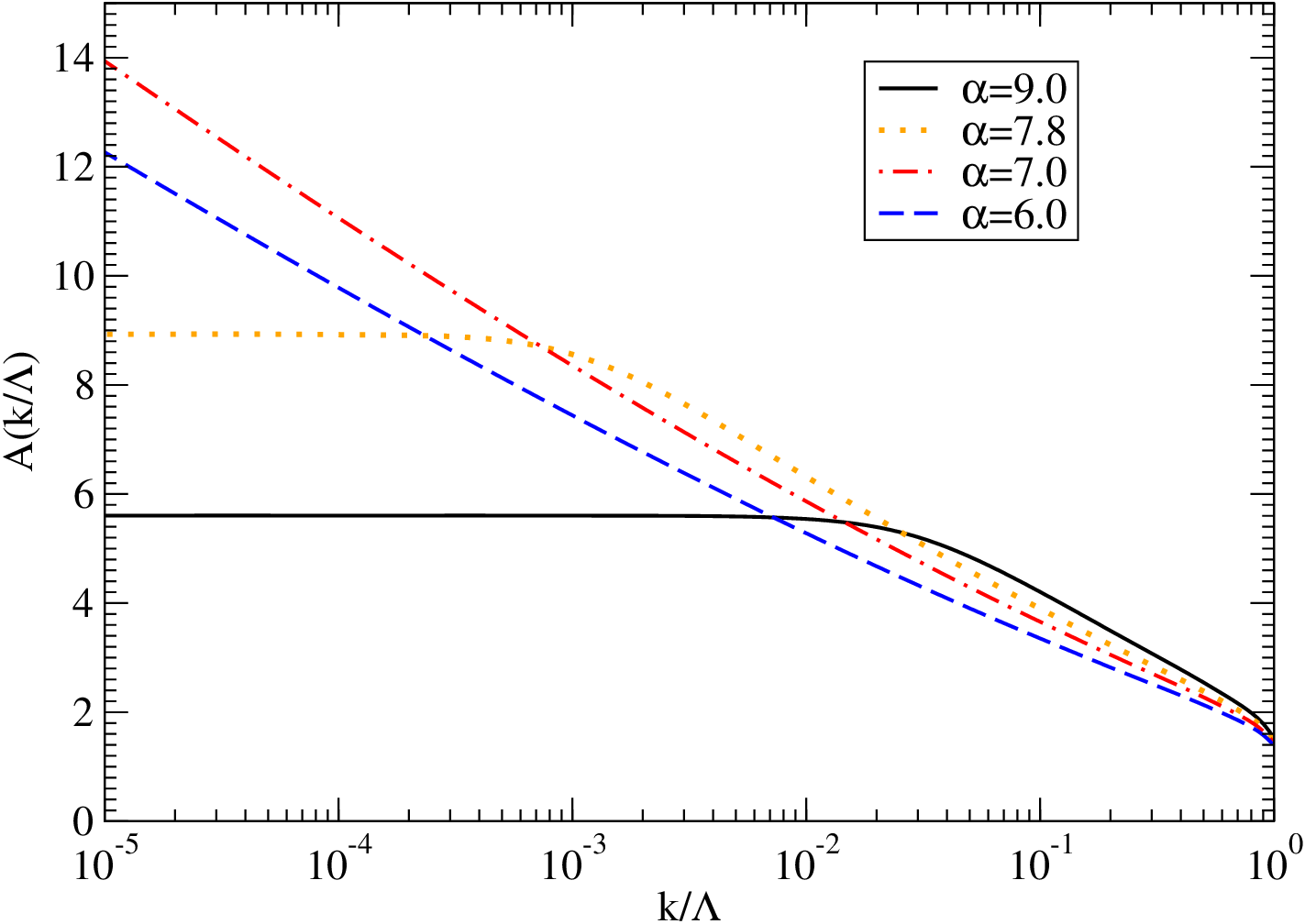}
\caption{\label{fig:A-B0}
Left: fermion mass function at zero momentum over the inverse coupling,
$1/\al$, together with two fits for values of the coupling close to the critical one,
see text for details.  
Right: momentum dependence of the Fermi velocity dressing function,
exemplified for the couplings $\al=6.0, 7.0, 7.8$ and $\al=9.0$, below and above the
corresponding critical coupling $\al_c \approx 7.65(5)$, respectively. Note the
logarithmic infrared divergence in the results for $\al=6.0, 7.0$, whereas for
$\al=7.8, 9.0$ this infrared divergence is screened by the dynamically generated
mass. All results are for $N_f=2$ here, see text for further details.}
\end{figure}

Having performed the bifurcation analysis which is appropriate precisely at
the critical coupling, we now return to the original system of coupled
(nonlinear) integral equations for the fermion propagator,
Eqs.~(\ref{eq:gapA}) and (\ref{eq:gapD}). To illustrate the potential effects
of a large $A$-function, here we focus on the frequency dependent RPA Coulomb
interaction with the one-loop photon polarization function of
\eq{one-loop}. Explicitly, the equations then read (with $z$ now as given in
(\ref{eq:Iz}), and $g= \alpha_\varepsilon N_f\pi/4$ as before):
\bea
A(p)=1+\frac{\alpha_\varepsilon}{2\pi^2}\frac{1}{p}\int_{0}^{\Lambda} dk\, k  \int_{0}^{\pi}
d\theta \frac{\cos \theta}{\sqrt{p^2+k^2-2\, p \,k \, \cos\theta}}
\frac{A(k)}{\sqrt{A^2(k)+\frac{\Delta^2(k)}{k^2 v_F^2}}}J(z,g)\, , 
\label{eq:gapA3}\\
\Delta(p)=m+ 
\frac{\alpha_\varepsilon}{2\pi^2}\int_{0}^{\Lambda} dk \int_{0}^{\pi}
d\theta \frac{1}{\sqrt{p^2+k^2-2\, p\, k\, \cos\theta}}
\frac{\Delta(k)}{\sqrt{A^2(k)+\frac{\Delta^2(k)}{k^2 v_F^2}}}J(z,g)\, ,
\label{eq:gapD3}
\eea
where we have also included a small bare mass $m$ via the tree-level
propagator in the gap equation which acts as an external field for explicitly
breaking the extended $U(2N_f)$ flavor symmetry of the low-energy theory down
to $U(N_f)\times U(N_f) $ as discussed below.

In the nonlinear case, the critical coupling is obtained from evaluating the
numerically determined mass function at zero momentum, in the chiral limit
$m=0$, which serves as an order parameter for chiral symmetry breaking together 
with an extrapolation towards the critical point using appropriate fit functions. 
Our numerical results together with the fits are shown in the left diagram of 
Fig.~\ref{fig:A-B0}. Numerically, we seem to find an exponential decrease of the 
order parameter close to the critical coupling indicating Miransky scaling 
similar to the case of ordinary QED$_3$ 
\cite{Miransky:1996pd,fial2004,Braun:2010qs}. Indeed, a corresponding fit of the Miransky 
type form
\begin{equation}
\Delta(0)/\Lambda = a_0 \exp\left(\frac{a_1}{\sqrt{\frac{1}{\alpha_c}-\frac{1}{\alpha}}}\right)
\end{equation}
delivers excellent results with the parameters $a_0 = 6.4 \times 10^{-4}$, $a_1 = -0.138$
and a critical coupling of $\alpha_c=7.68$. However, we also note that the fit form
\begin{equation}
\Delta(0)/\Lambda = \frac{|\alpha-\alpha_c|}{\left[\ln\left(\frac{a}{\alpha-\alpha_c}\right)\right]^3}
\end{equation}
extracted from analytical results using angular approximation \cite{Gusynin:unp}
works equally well with the parameter $a=345061$ and $\alpha_c=7.70$. Both values for
the critical coupling obtained in this way are within the range $\al_c\approx 7.65(5)$
obtained from bifurcation analysis in the last subsection. 

To demonstrate the effects of the dynamical mass generation, we show in the
right panel of \fig{fig:A-B0} our numerical results for the Fermi velocity
renormalization function at the couplings $\al=6$ and $\al=7$ as well 
as $\al=7.8$ and $\al=9$, below and above the critical coupling $\al_c$. 
We notice that in the symmetric phase the selfconsistent numerical solution
for $A(p)$ contains a logarithmic divergence at $p^2 \rightarrow 0$ as in the
GW approximation \cite{Popovici:2013fwp}. The slope of this logarithmic
increase grows with $\alpha $ until the logarithmic infrared divergence gets
suppressed due to the dynamical generation of a mass gap when the broken phase
is reached such that $A(p)$ approaches a finite value for $\al > \al_c$ in the
infrared. While a logarithmic behavior with further increasing slope persists
for a certain range of intermediate momenta also in the broken phase above
$\al_c$, the saturation value in the infrared decreases with $\al $ from there
on, and the momentum scale for this saturation hence increases. From an
experimental point of view, the function $A$ plays a very important role, as
the Fermi velocity enters the mass function, and many other graphene
observables. In fact, recent measurements \cite{elgo2012} provide evidence
that the spectrum of suspended graphene is indeed approximately logarithmic
instead of linear near charge-neutrality. This logarithmic increase
would eventually invalidate the instantaneous Coulomb approximation
for the electron interactions in graphene as pointed out in
\cite{Gonzalez94} already. Phenomenologically, on the other hand, the
smallest momenta reached in experiments are limited by the finite size of
realistic graphene sheets. For those one observes an increase of the
Fermi velocity by a factor (e.g.\ of the order of three in
\cite{elgo2012}). With our present results for the semimetal phase it
would typically take sheets that are larger by several orders of
magnitude to increase this logarithmic Fermi velocity renormalization 
factor from 3 to say 10. With the bare  $v_F/c \approx 1/300$  this
would still justify the instantaneous approximation with $Z=1$, as
explained below Eq.~(\ref{instphoton}), reasonably well.

\begin{figure}[t]
\vspace{0.5cm}
\centering
\includegraphics[width=0.48\linewidth]{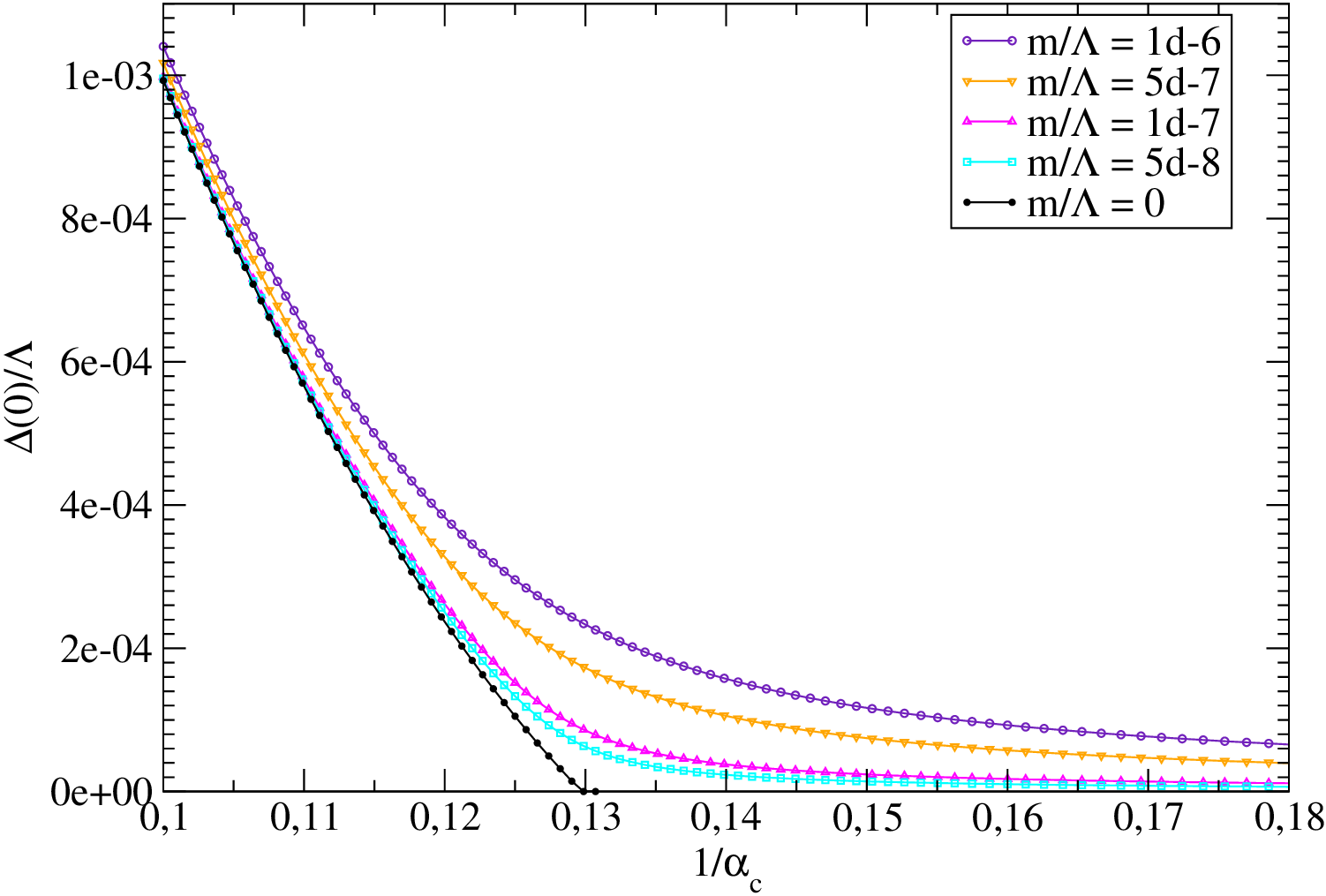}
\hfill
\includegraphics[width=0.48\linewidth]{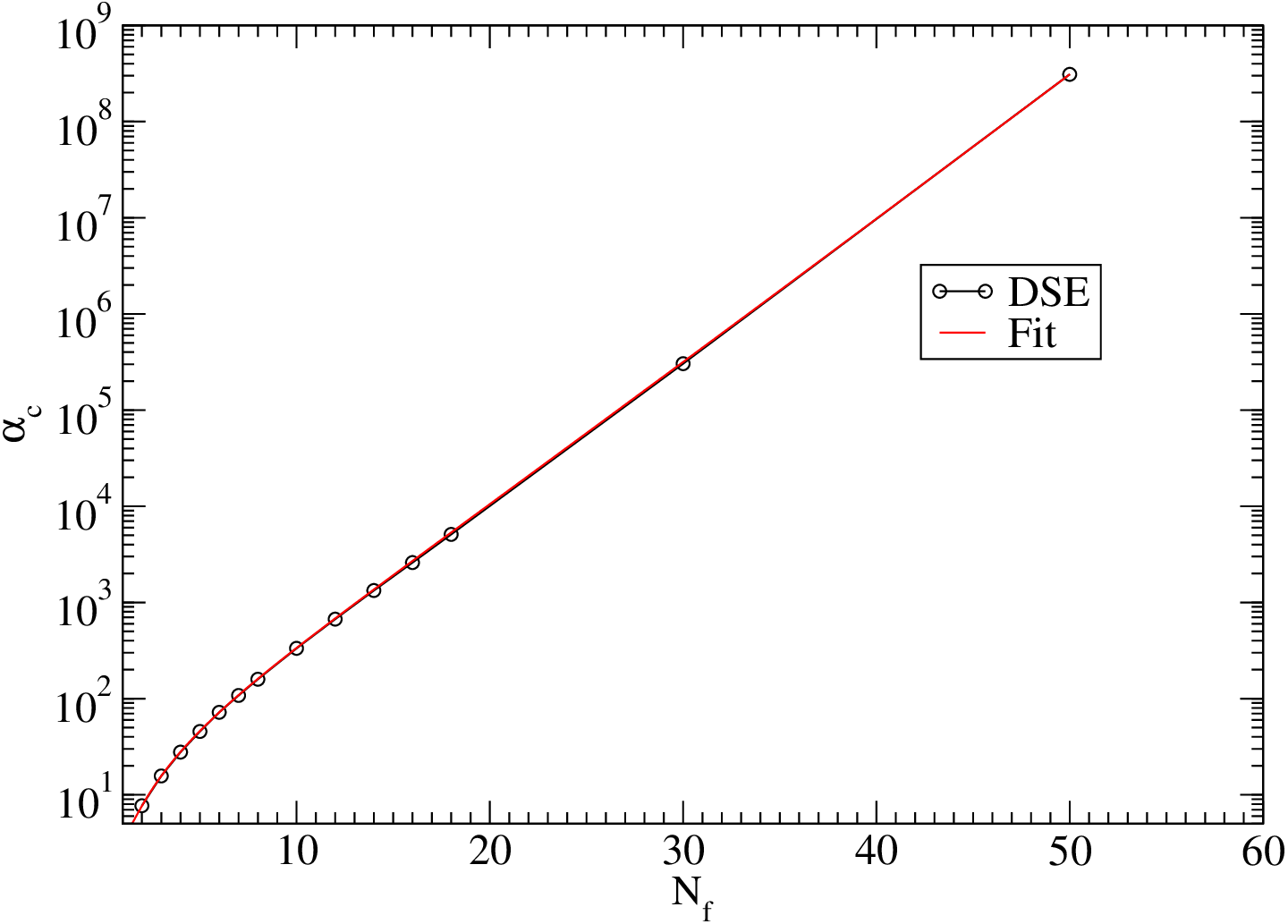}
\caption{\label{fig:mass}
Left: Mass dependence of the order parameter for $N_f=2$.  Right: Large $N_f$
dependence of the critical coupling.
}
\end{figure}

In order to study explicit symmetry breaking effects we have also determined
the zero-momentum limit of the mass function for a range of bare fermion
masses $m$. The corresponding results are plotted as a family of functions over
the inverse coupling in the left panel of Fig.~\ref{fig:mass}. The general
pattern is similar to the one seen in the lattice simulations
\cite{drula2009}.  The critical scaling is quite different, however. While
infinite volume extrapolations of lattice results have provided evidence of a
second order phase transition with associated critical exponents, here we
observe the typical Miransky scaling of QED$_3$ in the continuum approach as
mentioned above and discussed in more detail in
Ref.~\cite{Gamayun:2009em}. Consequently, the magnetic scaling of the order
parameter with the mass at the critical coupling is different from that in
\cite{drula2009}. Whether the infinite volume and chiral limit extrapolations
are reliable in such a case is not clear to us.
 
The explicit symmetry breaking by $m\not= 0$ could be due to a staggered
chemical potential with opposite signs on the two sublattices as induced by a
sublattice-dependent substrate, for example. It would then be a seed for
site-centered charge-density wave formation and as such break parity.  There
are other Dirac mass terms which lead to the same general breaking pattern of
the extended flavor symmetry, $ U(2N_f)\to U(N_f)\times U(N_f) $, but rather
correspond to bond-ordered phases \cite{Hou:2006qc}, however. The main
difference is whether the residual $ U(N_f)\times U(N_f) $ mixes the two Dirac
points or not \cite{gush2007,Mesterhazy:2012ei}. This can not be distinguished
on the level of the gap equation because it depends on the particular choice
of the basis used for the reducible four-dimensional representation of the
$\gamma $-matrices which we did not have to specify in the derivation of
\eq{eq:gapD3}.\footnote{Yet another realization of the same symmetry breaking
  pattern is possible when the sign of the mass term is made spin-dependent as
  it is actually done in lattice simulations to avoid the sign problem
  \cite{Ulybyshev:2013swa}. It then acts as an external field for
  anti-ferromagnetic order. To describe this we would have to treat the
  $N_f=2$ flavors separately here.}

Finally, let us shortly comment on the dependence of the critical coupling on
the number of flavors $N_f$.  The corresponding curve displayed in the right
panel of Fig.~\ref{fig:mass} can be fitted with an expression of the form
\be
f(N_f)=12.7\, e^{\frac{N_f}{3}\sqrt{\frac{N_{fc}}{N_{fc}-N_{f}}}}-14.4\, N_{f}^{0.25}
\ee
from which we may read off a value for the critical number of (pseudo)fermion
species of about $N_{fc}\sim 1200$ for which the critical coupling
diverges. While this particular number is certainly not reliable for the
reasons discussed above, the general finding of the very existence of a finite
value for $N_{fc}$ would confirm a conjecture of Son in
Ref.~\cite{Son2007}. It remains to be seen, however, whether this result
survives a selfconsistent treatment of the particle-hole polarization
function.

\section{Summary and outlook}\label{sec5}

In this study we have determined critical couplings for the chiral phase
transition from the fermion Dyson-Schwinger equation (DSE) with long-range
Coulomb interactions in a low-energy model for monolayer graphene at half
filling. As compared to previous DSE studies of this model we have for the
first time dynamically included a non-trivial Fermi velocity dressing function
in our selfconsistent solutions for the fermion propagator.  We have thereby
compared the effects of static as well as fully frequency-dependent Lindhard
screening with the bare Coulomb interaction. In all three cases, the
selfconsistent inclusion of the non-trivial Fermi velocity dressing function
had dramatic effects, indicating that a substantial renormalization of the
Fermi velocity should occur at strong coupling in agreement with experimental
evidence from cyclotron-mass measurements in suspended graphene
\cite{elgo2012}.  At the same time, such large Fermi velocity renormalizations
also indicate that RPA Coulomb interactions with one-loop polarization
function considerably overestimate the screening effects. Whether the
screening is nevertheless strong enough for suspended graphene to remain in
the semimetal phase remains to be seen from a fully dynamical inclusion of the
particle-hole polarization in a simultaneous solution of both, the fermion and
the Coulomb photon DSEs in the future.

Including the running Fermi velocity renormalization function, here we
obtained $\al_c \approx 1.22$ for the critical coupling of the
semimetal-insulator transition without screening as compared to $\al_c \approx
7.7$ with the fully frequency-dependent Lindhard screening in the RPA Coulomb
interaction. We have argued that a selfconsistent treatment of the
particle-hole polarization should lead to a critical coupling between these
two extremes with an expected tendency towards values closer to the lower
bound. This leaves open the possibility that the critical coupling is larger
than the bare coupling $\al_0=2.16$ for suspended graphene. Quite likely,
however, additional screening from the electron states in the $\sigma $ bands
of graphene might have to be included in a more realistic calculation to
achieve this \cite{Wehling:2011}.

Although our study is indicative, a number of caveats remain.  First of all we
need to include the particle-hole polarization effects dynamically and
selfconsistently to study their effect on $\al_c$. When comparing with lattice
results it will also be important to take finite volume effects into
account. Finally, when comparing with experiment, other type of interactions
such as four-fermi couplings might also be important and need to be included
in the model. First important steps in this direction have been discussed in
\cite{Gamayun:2009em,Mesterhazy:2012ei}.

\vspace{.6cm}

\leftline{\bf Acknowledgments}

\vspace{.2cm}
We are grateful to V.~P.~Gusynin for pointing out numerical inaccuracies in a previous 
version of the manuscript and for sending us his private notes on analytical
solutions of some of the equations.
We would like to thank P.~Buividovich, D.~Mesterhazy, M.~Ulybyshev, D.~Smith,
S.~Strauss and the late M.~Polikarpov for helpful discussions.  This work was
supported by the Deutsche Forschungsgemeinschaft within SFB 634, by the
Helmholtz Young Investigator Group No.~VH-NG-332, and by the European
Commission, FP7-PEOPLE-2009-RG, No. 249203.


\bibliography{biblio}

\end{document}